Research article

# Clustering of molecular dynamics trajectories *via* peak-picking in multidimensional PCA-derived distributions.

Athanasios S. Baltzis,  Panagiotis I. Koukos  & Nicholas M. Glykos*

*Department of Molecular Biology and Genetics, Democritus University of Thrace, University campus, 68100 Alexandroupolis, Greece, Tel +30-25510-30620, Fax +30-25510-30620, http://utopia.duth.gr/glykos/ , glykos@mbg.duth.gr*

# Abstract

We describe a robust, fast, and memory-efficient procedure that can cluster millions of structures derived from molecular dynamics simulations. The essence of the method is based on a peak-picking algorithm applied to three- and five-dimensional distributions of the principal components derived from the trajectory and automatically supports both Cartesian and dihedral PCA-based clustering. The density threshold required for identifying isolated peaks (which correspond to discrete clusters) is determined through the application of a variance-explained criterion which allows for a completely automated clustering procedure with no user intervention. In this communication we describe the algorithm and present some of the results obtained from the application of the method as implemented in the molecular dynamics analysis programs `carma`, `grcarma`. and `cluster5D`. We conclude with a discussion of the limitations and possible pitfalls of this method.

# Keywords

# 1. Introduction

Meaningful clustering of macromolecular molecular dynamics trajectories is a complex and open-ended question. The main reason for this complexity is that the mode of clustering depends on the aim of the analysis performed and is not a fundamental property of any given trajectory. For example, significantly different clustering procedures are needed if the aim of the analysis is to identify putative transition state ensembles from a folding molecular dynamics simulation, as opposed to identifying, for example, the most prominent and stable molecular conformations from this same folding trajectory. What this discussion implies, then, is that the development of a universally optimal clustering algorithm for macromolecular simulation is highly unlikely -if not meaningless. The conclusion from this analysis is clear : any new clustering algorithm -especially those that have been fully automated- must explicitly define the aims and limitations of the clustering performed.

Here we describe a PCA-based peak-picking algorithm which has been implemented in the molecular dynamics analysis programs `carma`[1], `grcarma`[2] and `cluster5D`. The main reason that led us to develop this algorithm was mostly technical : the great majority of clustering procedures currently available have very significant physical memory requirements and are computationally quite expensive which makes their application to molecular dynamics trajectories containing several millions of structures rather impractical. Additionally, for most of these methods extra steps of pre- and post-processing of the trajectories are needed which further limits their direct application to the primary simulation data. It is for this reason that we have developed the algorithm described in this communication which has been shown to be able to cluster millions of frames both extremely fast and with negligible physical memory requirements. As will be discussed later, the aim of this clustering procedure is to quickly identify prominent molecular configurations based on a principal component analysis of the trajectories and supporting both Cartesian and dihedral PCA.

In the paragraphs that follow we describe the algorithm, present results from its applications to several different macromolecular trajectories, and then critically discuss its limitations for the analysis of biomolecular trajectories.

# 2. Methods

## 2.1 Algorithms

The algorithm is essentially based on identifying isolated peaks in a three- or five-dimensional density map obtained from the distribution of the trajectory's principal components. In more detail.

1. The trajectory is analyzed using either Cartesian or dihedral PCA [1,2 and references therein]. The result is a matrix containing the values of the top three (or five) principal components for each and every structure recorded in the trajectory. The choice of the maximum number of supported dimensions (five) is arbitrary and was selected in order to make the programs useable even on computing machines with limited hardware resources.

2. The values of the principal components obtained from step (1) above are converted to a three- or five-dimensional density distribution map. In these maps, the three (or five) axes correspond to the respective principal components, and the value of each point of the map is equal to the number of structures from the trajectory that have PC values closest to the that point. Figure 1 shows a two-dimensional example aiming to exemplify this procedure : The two axes in panel (A) of this Figure correspond to the top two principal components (the first component horizontal, the second vertical). The density of the map itself (with hot colors indicating high density) is proportional to the number of structures (from the trajectory) that have values for their principal components corresponding to the specific point of the PC plane.

3. The initial three/five dimensional map obtained in step (2) above is smoothed using nearest-neighbor averaging. The aim of this smoothing step is to reduce the amount of statistical noise in the maps which arises from the limited and discontinuous sampling of the trajectory. It should be noted that even after this smoothing step, the maps still contain a significant amount of noise as can be seen in Figure 1(B) which depicts the logarithm of the distribution shown in panel (A).

4. Having obtained the smoothed maps, the aim is to identify isolated peaks in these distributions that have a density higher than a given threshold. The choice of threshold for identifying isolated peaks is critical in the sense that an erroneously high threshold will lead to loss of structural information while an erroneously low threshold will fail to discriminate between distinct molecular conformations. We have implemented an automatic procedure for calculating a reasonable value for the density threshold through the application of a variance-explained criterion. The principal idea is that we test a large number of different thresholds (starting from a very high value and progressively lowering it until we reach the map's mean density) and for each of these different thresholds we calculate the percentage of the map's variance that is explained by the peaks above the threshold. Our implementation aims for a threshold that can explain at least 80% of the original PC map's variance.

5. In the last step, all peaks above the selected threshold are identified and classified, with each of these peaks corresponding to a distinct cluster. The trajectory frames belonging to each cluster can easily be identified (and reported) via a second pass through the principal component matrix described in step (1) above. Panel (C) in Figure 1 shows the results obtained by applying the method to the distribution shown in panel (A) of the same figure. Notice how some of the less prominent conformations (lower density peaks) escaped detection. This and other limitations of the method will be discussed in the Discussion section.

## 2.2 Implementation and availability

The algorithm described above has been fully implemented (for both the three- and five-dimensional cases) in the free, open-source programs `carma`, `grcarma` and `cluster5D` which are immediately available for download *via* standard repositories. Complete packages containing pre-compiled executables suitable for all major architectures are also available via http://utopia.duth.gr/glykos/Carma.html .

# 3. Results

The method described in the previous sections has been tested with numerous trajectories available to us and covering everything from peptides[3-8] to large proteins[9,10] and from extreme stability[3,6,7] to significant disorder[4,5,8,10,18]. Since the initial release of `carma`, several groups from the molecular dynamics community have tested and applied this algorithm. To our knowledge, and within the limitations discussed in the next section, the method has been shown to be robust, extremely fast, and capable of clustering millions of frames with minimal hardware requirements. Case studies exemplifying the application of the algorithm in real problems can be found in several papers[3-17].

# 4. Discussion

The major limitation of the method described above is that it does not even attempt to cluster all structures from the trajectory : the aim of the algorithm is to efficiently identify the most prominent molecular conformations, and not to comprehensively assign each frame of a trajectory to a cluster. Additionally, the fixed highest dimensionality of the PC-derived maps may in some cases lead to loss of structural information (this is more probable with highly flexible/disordered systems where a larger number of principal components may be required to capture the structural variance). Figure 2 demonstrates some of the limitations of the method using a hypothetical one-dimensional distribution : Depending on the peak-picking threshold selected by the variance-explained criterion, low lying peaks may escape detection (cluster D in Figure 2) or closely related -but otherwise distinct- conformations may be assigned to the same cluster (clusters A and B). It could be argued that a better (gradient-based) algorithmic treatment could be devised that would alleviate both of these problems. Practical experience with the algorithm has shown, however, that the presence of significant noise in the primary data, together with the requirement to be able to cluster even trajectories containing only hundreds (instead of millions) of frames severely limits the ability to meaningfully extend the procedure described above. Having noted these limitations, we should close by stating that from the numerous simulations that this algorithm has been applied we have never observed a case where this method failed to produce a reasonable and convincing initial characterization of a trajectory's most prominent molecular configurations.

# Figure Captions

**Figure 1 : Schematic representation of the peak-picking algorithm.** Panel A shows a pseudo-color representation of the distribution of the top two principal components obtained from a 2.1 μs molecular dynamics simulation of hepta-Alanine. Panel B is the same distribution on a logarithmic scale. Panel C indicates the five clusters automatically identified by the algorithm. Note that the actual clustering algorithm operates in either the three or five dimensional PCA space, and, thus, what is shown in this figure is a lower dimensionality projection.

**Figure 2 : Artifacts and limitations of the peak-picking algorithm.** This diagram shows a hypothetical density distribution for one principal component. The horizontal axis is the value of the principal component, the vertical axis is the respective density (number of structures with the corresponding value of the principal component). The red horizontal line is a hypothetical density threshold for identifying peaks which was selected through the variance-explained criterion. Notice how (a) only a subset of structures from the trajectory are assigned to clusters, (b) that some clusters escape detection (cluster D in the diagram), and, (c) how closely related clusters are being treated as a single peak (clusters A and B in the diagram).

# Figure 1

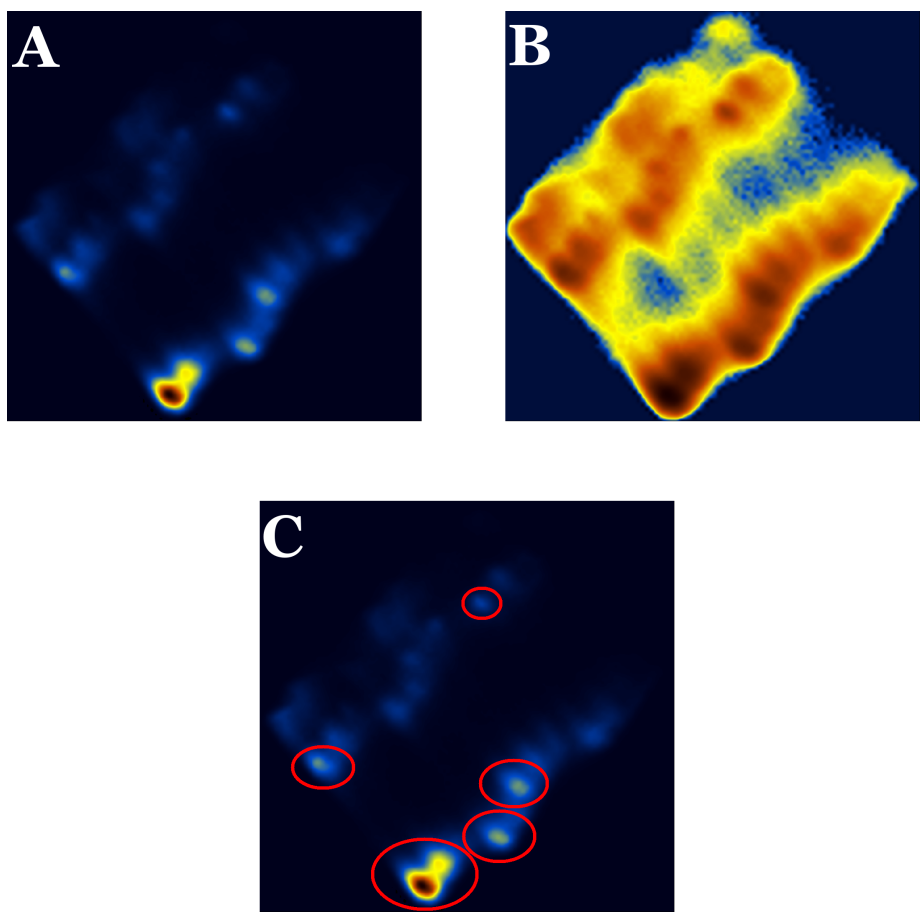

**Figure 1 : Schematic representation of the peak-picking algorithm.** Panel A shows a pseudo-color representation of the distribution of the top two principal components obtained from a 2.1 μs molecular dynamics simulation of hepta-Alanine. Panel B is the same distribution on a logarithmic scale. Panel C indicates the five clusters automatically identified by the algorithm. Note that the actual clustering algorithm operates in either the three or five dimensional PCA space, and, thus, what is shown in this figure is a lower dimensionality projection.

# Figure 2

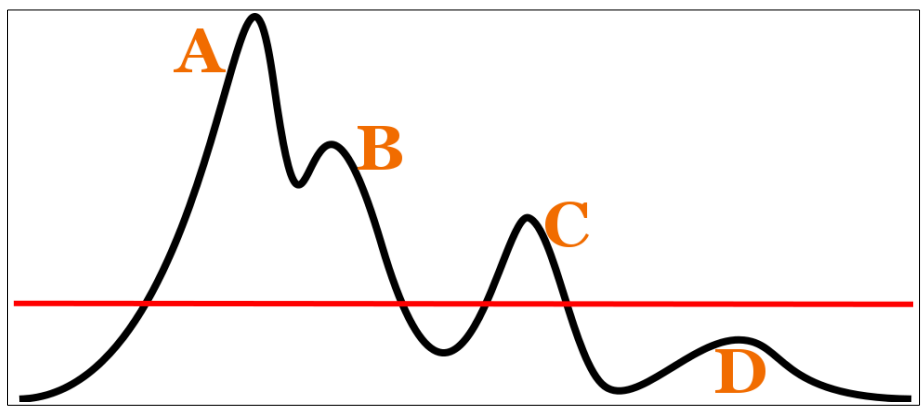

**Figure 2 : Artifacts and limitations of the peak-picking algorithm.** This diagram shows a hypothetical density distribution for one principal component. The horizontal axis is the value of the principal component, the vertical axis is the respective density (number of structures with the corresponding value of the principal component). The red horizontal line is a hypothetical density threshold for identifying peaks which was selected through the variance-explained criterion. Notice how (a) only a subset of structures from the trajectory are assigned to clusters, (b) that some clusters escape detection (cluster D in the diagram), and, (c) how closely related clusters are being treated as a single peak (clusters A and B in the diagram).